\documentclass[aps,showpacs,tightenlines,preprint]{revtex4}
\usepackage{amsmath}
\usepackage{epsfig}
\newcommand{\bra}{\langle}
\newcommand{\ket}{\rangle}
\newcommand{\sixj}[6]{
     \left\{ \begin{array}{ccc}
              #1 & #2 & #3 \\
              #4 & #5 & #6 
            \end{array}  \right\} } 
\newcommand{\ratio}[2]{{\textstyle \frac{#1}{#2}}}
\newcommand{\bs}[1]{\ensuremath{\boldsymbol{#1}}}

\begin{document}
\title{Four-quark spectroscopy within the hyperspherical formalism}

\author{N. Barnea$^1$, J. Vijande$^2$, A. Valcarce$^2$}
\affiliation{
$^{1}$ The Racah Institute of Physics, The Hebrew University, 91904, Jerusalem, Israel\\
$^{2}$ Grupo de F\'{\i}sica Nuclear and IUFFyM, Universidad de Salamanca,
E-37008 Salamanca, Spain}
\date{\today}

\begin{abstract}
We present a generalization of the hyperspherical harmonic formalism to study
systems made of quarks and antiquarks of the same flavor. This generalization
is based on the symmetrization of the $N-$body wave function 
with respect to the symmetric group using the Barnea and Novoselsky algorithm.
The formalism is applied to study four-quark systems by means of a constituent
quark model successful in the description of the two- and three-quark systems.
The results are compared to those obtained by means of variational approaches.
Our analysis shows that four-quark systems with exotic $0^{+-}$ and non-exotic
$2^{++}$ quantum numbers may be bound independently of the mass of the quark.
$2^{+-}$ and $1^{+-}$ states become attractive only for larger mass of the quarks.
\end{abstract}

\bigskip

\maketitle
\section{Introduction}

The understanding of few-body systems relies in our capability to design
methods for finding an exact or approximate solution of the $N-$body problem.
In two-, three-, and four-body problems it is possible to obtain mathematically
correct and computationally tractable equations such as the Lippmann-Schwinger,
Faddeev and Yakubovsky equations describing exactly, for any assumed
interaction between the particles, the motion of few-body systems \cite{Bel90}.
However, the exact solution of these equations requires sophisticated
techniques whose difficulty increases when increasing the number of particles.

There are a countless number of examples of quantum-mechanical few-body
systems, from few-electron quantum dots in solid state physics to constituent
quarks in subnuclear physics. The intricate feature of the few-body systems is
that they develop individual characters depending on the number of constituent
particles. The most important cause of these differences are the correlated
motion and the Pauli principle. This individuality requires specific methods
for the solution of the few-body Schr\"odinger equation, approximate solutions
assuming restricted model spaces (as for example the mean field approximation)
failing to describe these systems.

The solution of any few-particle system may be found in a simple and unified
approach. A recent widely used method is the stochastic variational
\cite{Suz98}, a variational approach where the trial wave function is generated
by a random search on an adequate basis. An alternative powerful tool is an
expansion of the trial wave function in terms of hyperspherical harmonic (HH)
functions. The idea is to generalize the simplicity of the spherical harmonic
expansion for the angular functions of a single particle motion to a system of
particles by introducing a global length $\rho$, called the hyperradius, and a
set of angles, $\Omega$. For the HH expansion method to be practical, the
evaluation of the potential energy matrix elements must be feasible. The main
difficulty of this method is to construct HH functions of proper symmetry for a
system of identical particles. This is a difficult problem that may be
overcome by means of the HH formalism based on the symmetrization of the
$N-$body wave function with respect to the symmetric group using the Barnea and 
Novoselsky algorithm \cite{Nir9798}. This method, applied in nuclear physics
for $N\leq 7$ \cite{Bac04}, has only been applied to quark physics for $N=3$ \cite{Vij05}.
Therefore, its generalization would be ideally suited for the study of the
properties of multiquark systems.

During the last few years there has been a renewed interest on the possible
existence of multiquark states, specially four- (two quarks and two antiquarks)
and five-quark (four quarks and one antiquark), in the low-energy hadron
spectroscopy. Theoretically, the possible existence of four-quark bound
states was already suggested thirty years ago, both in the light-quark
sector by Jaffe \cite{Jaf77} and in the heavy-quark sector by Iwasaki
\cite{Iwa76}. Experimentally, there were several analysis suggesting the
existence of non-$q\bar q$ states in the high energy part of the
charmonium spectrum \cite{Ros76}. The recent series of discoveries of
new meson resonances whose properties do not fit into the predictions of the
naive quark model, has reopened the interest on the possible role
played by non-$q\bar q$ configurations in the hadron 
spectra \cite{Jaf04,Ams04}. Among them one could mention the
$X(3872)$ \cite{Cho03}, the $Y(4260)$ \cite{Aub05} or the
new $D$ and $D_s$ resonances \cite{Aub03}. Out of the
several interpretations proposed for these states, those
based on four-quark states are majority \cite{Bar05}. 
There seems to be a consensus that a four-quark system
containing two-light and two-heavy quarks, $qq \overline Q \overline Q$, 
is stable against dissociation into two mesons, $q\overline Q$,
if the ratio of the mass of the heavy to the light quark
is large enough \cite{Ade82,Hel87}. While the conclusions 
agree for the $qq\bar b \bar b$ system, there are more
discrepancies about the $qq\bar c \bar c$ 
system \cite{Car88,Sil93,Vij03c}.

The possible existence of four-quark states in the low-lying meson 
spectra has been used to advance in the understanding of the
light-scalar mesons, $J^{PC}=0^{++}$. In this case,
four-quark states have been justified to coexist with $q\bar q$ states in the energy region
below 2 GeV because they can couple to $0^{++}$ without
orbital excitation \cite{Jaf77}, and therefore conventional $q \bar q$ states
are expected to mix with four-quark states to yield
physical mesons \cite{Ams04,Vij03b}.

The four-heavy quark states have not received as much attention
in the last years as the ones containing light quarks. 
Iwasaki \cite{Iwa76}, based on the constituent quark mass value obtained within a 
string model for hadrons, argued that four-charm quark states could exist in the 
6 GeV energy region. Using a variational method with gaussian trial radial wave functions and 
two-body potentials based on the exchange of color
octets between quarks, Ader {\it et al.} \cite{Ade82} obtained non stable 
$QQ\overline Q\overline Q$ states. However, using
a potential derived from the MIT bag model in the
Born-Oppenheimer approximation the same authors concluded that the $cc\bar c\bar c$ state
was bound by 35 MeV. Heller and Tjon \cite{Hel87} considered also the 
MIT bag model improving the Born-Oppenheimer approximation
with a correct treatment of the kinetic energy, non stable 
$QQ\overline Q\overline Q$ states were found.
In Ref. \cite{Sil92} $L=0$ $QQ\overline Q\overline Q$ states were analyzed
in the framework of a chromomagnetic model where only a constant hyperfine potential,
not depending on radial coordinates, was retained. With these
assumptions no bound $QQ \overline Q\overline Q$ states were found. 
A similar conclusion was obtained in Ref. \cite{Sil93} using the 
Bhaduri potential and solving variationally the four-body problem in a harmonic 
oscillator basis. Recently, Lloyd and Vary \cite{Llo04} 
investigated the $cc\bar c\bar c$ system
using a nonrelativistic hamiltonian inspired by the one-gluon exchange
potential diagonalizing the hamiltonian in a harmonic oscillator basis, 
obtaining several close-lying bound states.

The discussion above illustrates that the theoretical predictions for the
existence of four-quark systems differ depending basically on the method used to solve
the four-body problem and the interaction employed. It is our aim 
in this work to make a general study of four-quark systems of identical flavor
in an exact way. For this purpose we will generalize the HH method
\cite{Bar00}, widely used in traditional nuclear physics for the study of
few-body nuclei, to study four-quark systems. There are two main difficulties,
first the simultaneous treatment of particles and antiparticles, and second
the additional color degree of freedom. As a test of our formalism we will
recover some results present in the literature. Finally, we will make a general
study of four-quark systems of identical flavor by means of the constituent
quark model of Ref. \cite{Vij03} that provides with a realistic framework
describing in a correct way the general aspects of the meson and baryon
spectra.

The paper is organized as follows. In the next section the formalism
necessary to build a color singlet wave function with well-defined parity and
$C-$parity quantum numbers is described and discussed. In Sect. 
\ref{model} we introduce the general
features of the constituent quark model used. In 
Sect. \ref{results}  we present and analyze the results 
obtained for the four-quark systems. Finally, in 
Sect. \ref{summary} we resume our most
important conclusions.

\section{General formulation of the problem}
\label{secII}

The system of two quarks and two antiquarks with the same flavor can be regarded
as a system of four identical particles. Each particle carries
a $SU(2)$ spin label and a $SU(3)$ color label. Both quarks and antiquarks are
spin $\ratio{1}{2}$ particles, but whereas a quark color state
belongs to the $SU(3)$ fundamental representation 
$[3]$, an antiquark color state is a member of the fundamental 
representation $[\bar 3]$.   
The four-body wave function is a sum of outer products
of color, spin and configuration terms
\begin{equation}
    | \phi \ket = |{\rm Color}\ket |{\rm Spin}\ket | R \ket
\end{equation}
coupled to yield an antisymmetric wave function with
a set of quantum numbers that reflects the symmetries of the system.
These are the total angular momentum quantum number $J$, its projection $J^z$, 
and the $SU(3)$ color
state $G$ (to avoid confusion with the charge conjugation quantum number the
color degree of freedom will be labeled as $G$ from now on), which by assumption must belong for physical states
to the $SU(3)$ color singlet representation. 
Since QCD preserves parity, parity is also a good 
quantum number. Another relevant quantum number to the system
under consideration it is the $C-$parity, $C$, 
i.e., the symmetry under interchange of 
quarks and antiquarks.

To obtain a solution of the four-body Schr\"odinger equation we 
eliminate the center of mass and use the relative, Jacobi,
coordinates $\vec{\eta}_1,\vec{\eta}_2,\ldots,\vec{\eta}_{A-1}$. Then
we expand the spatial part of the 
wave-function using the HH basis. In this formalism the Jacobi coordinates 
are replaced by one radial coordinate,  
the hyperradius $\rho$, and a set of ($3A-4$) angular coordinates 
$\Omega_A$. The HH basis functions are eigenfunctions of the hyperspherical
part of the Laplace operator.
An antisymmetric $A$--body basis functions 
with total angular momentum $J_A,J^z_A$, color $G_A $
and $C$-parity $C$, are given by,
\begin{eqnarray} \label{HH_A}
| n {K}_A J_A J^z_A G_A C \Gamma_A \alpha_A \beta_A \ket  & = &
\cr & & \hspace{-25mm}
      \sum_{Y_{A-1}}
      \frac{\Lambda_{\Gamma_{A},Y_{A-1}}}{\sqrt{| \Gamma_{A}|}} \,
      \left[ | K_A L_A M_A \Gamma_A Y_{A-1} \alpha_A \ket
             | S_A S^z_A G_A C 
               \, \widetilde{\Gamma}_{A},\widetilde{Y}_{A-1}
                  \, \beta_A \ket 
          \right]^{J_A J^z_A} | n \ket \,,
\end{eqnarray}
where
\begin{equation}
  \bra \rho | n \ket \equiv R_n(\rho)
\end{equation}
are the hyperradial basis functions, taken to be Laguerre functions.
\begin{equation}
\bra \Omega_A | K_A L_A M_A \Gamma_A Y_{A-1} \alpha_A \ket
      \equiv 
{\cal Y}^{[A]}_{K_A L_A M_A \Gamma_A Y_{A-1} \alpha_A}(\Omega_A)   
\end{equation}
are HH functions with hyperspherical angular momentum $K=K_A$, 
and orbital angular momentum quantum numbers ($L_A, M_A$) that belong
to well-defined irreducible representations (irreps) 
$\Gamma_{1} \in \Gamma_2 \ldots \in \Gamma_A $ of the permutation 
group--subgroup chain  
${\cal S}_1 \subset {\cal S}_2 \ldots \subset {\cal S}_A $,
denoted by the Yamanouchi symbol 
$[ \Gamma_A, Y_{A-1} ] \equiv [ \Gamma_A,\Gamma_{A-1},\ldots,\Gamma_1 ]$.
The dimension of the irrep $\Gamma_{m}$ is denoted by $| \Gamma_{m}|$
and $\Lambda_{\Gamma_{A},Y_{A-1}}$ is a phase factor \cite{NKG88}.
Similarly, the functions 
\begin{equation}
   \bra s^z_1..s^z_A, g_1..g_A 
   |  S_A S^z_A G_A C \, \widetilde{\Gamma}_{A},\widetilde{Y}_{A-1} 
                                                      \beta_A \ket 
   \equiv 
   \chi^{[A]}_{S_A S^z_A G_A \, \widetilde{\Gamma}_{A},\widetilde{Y}_{A-1}
                  \, \beta_A}(s^z_1..s^z_A, g_1..g_A)
\end{equation}
are the symmetrized color--spin basis functions, given in terms of the
spin projections ($s^z_i$) and color states ($g_i$) of the particles.
The quantum numbers $\alpha_A$, and $\beta_A$ are used to remove the degeneracy 
of the HH and color--spin states, respectively. For the construction
of the symmetrized HH basis we shall use the algorithm of Barnea and 
Novoselsky \cite{Nir9798}, which utilizes the group of kinematic rotations.
For the color--spin subspace, we will present in the following 
subsections a method to transform the standard basis
into a symmetrized color--spin basis with well defined color and $C-$parity.
The calculation of the Hamiltonian matrix-elements between the 
antisymmetric basis functions, Eq. (\ref{HH_A}), is practically the same
as in the nuclear physics case, replacing isospin by color, and the reader
is referred to Barnea {\it et. al.} \cite{BLO99}.

\subsection{The construction of symmetrized color-spin states}
The starting point for the construction of color-spin states with 
well-defined permutational symmetry, color and spin quantum numbers,
are the single particle states. The spin state of the $i$'th particle, quark or antiquark,
is given by $|s=\ratio{1}{2} \; s^{z}=\pm \ratio{1}{2} \ket $.
Using the Clebsh-Gordan coefficients one can successively couple the spin states
to construct an $A$-body state with well defined total spin 
\begin{eqnarray}  
 | S_A S^z_S S_{A-1} S_{A-2} \ldots S_1 \ket  & = &
 [[[|\ratio{1}{2}\ket|\ratio{1}{2}\ket ]^{S_2} 
    |\ratio{1}{2}\ket ]^{S_3} \ldots 
    |\ratio{1}{2}\ket ]^{S_A S^z_A}\,.
\end{eqnarray}
Theoretically one can adopt the same procedure for the color states replacing
the $SU(2)$ Clebsh-Gordan coefficients with the appropriate $SU(3)$ ones.
In practice, however, the $SU(3)$ coefficients are more involved
and we shall therefore follow a different path while dealing with color.
Each quark (antiquark) color state is one of the three states of the triplet
representation $[3]$ ($[\bar 3]$) of $SU(3)$. These states are eigenstates
of the operators 
($\mathbf{h}_1,\mathbf{h}_2$) which form the Cartan subalgebra of $SU(3)$.
Let us denote by $ | g_i \ket$ the color state of the $i$'th
particle, $| g_i \ket={|1\ket, | 2 \ket, | 3 \ket}$ for quark states
and ${|\bar1\ket, | \bar2 \ket, | \bar3 \ket}$ for antiquark states,
\begin{equation}
 \mathbf{h}_1 | g_i \ket = g^1_i | g_i \ket \, ; \,
 \mathbf{h}_2 | g_i \ket = g^2_i | g_i \ket \,.
\end{equation}
The eigenvalues $(g^1_i,g^2_i)$ (weights) uniquely label the quark (antiquark) states,
in particular we can write
\begin{eqnarray}\label{su3_33bar_states}
 |     1 \ket = & |\textstyle(+\frac{1}{2},+\frac{1}{3})\ket & ; \;
 |\bar 1 \ket =   |\textstyle(-\frac{1}{2},-\frac{1}{3})\ket \cr
 |     2 \ket = & |\textstyle(-\frac{1}{2},+\frac{1}{3})\ket & ; \;
 |\bar 2 \ket =   |\textstyle(+\frac{1}{2},-\frac{1}{3})\ket \cr
 |     3 \ket = & |\textstyle(     0      ,-\frac{2}{3})\ket & ; \;
 |\bar 3 \ket =   |\textstyle(     0      ,+\frac{2}{3})\ket \,.
\end{eqnarray}
An $A$-body color state is a product of the single particle color states.
This state does not belong to a well-defined irreducible representation of
the color $SU(3)$ group, but instead it
 is characterized by the ``color projections''
$G^1_A=g^1_1+g^1_2+\ldots g^1_A$ and $G^2_A=g^2_1+g^2_2+\ldots g^2_A$,
very much the same way as the ``m-scheme'' states used in nuclear structure 
calculations which have $J^z$ as a good quantum number but not $J$. 
From now on we shall use the notation $G_A$ for the pair of eigenvalues
$(G^1_A,G^2_A)$.
The combined color-spin single particle state is 
$|s_i=\ratio{1}{2} s^z_i g_i \ket$, and the $n$-particle
states are given by
\begin{equation} \label{chi_cs}
 | \chi_{GS} \ket = [[[| \ratio{1}{2} g_1\ket| \ratio{1}{2} g_2\ket]^{S_2} 
                       | \ratio{1}{2} g_3\ket]^{S_3} \ldots
                       | \ratio{1}{2} g_n\ket]^{S_n S^z_n} \, .
\end{equation}
In the following we shall present a method to transform these states into basis
states with well-defined permutational symmetry, color and
$C-$parity. This method is based on an algorithm by Novoselsky, Katriel and
Gilmore (NKG) \cite{NKG88} who devised a recursive method 
to split an invariant vector
space into states with well-defined permutational symmetry.
By now, this algorithm has been used to construct symmetrized states in
broad range of problems, and therefore we shall not 
give a detailed account
of the method but rather outline the main ideas. For a clear description
of the method the interested reader is referred to \cite{Akiva89a}.

The main tool of the NKG algorithm is the transposition class-sum operator.
For the $n$-particle permutation group ${\cal S}_n$, this operator
is the sum over all possible transpositions,
\begin{equation}
 \bs{\cal C}_2[{\cal S}_n]=\sum_{i<j}^n (i,j)  \;.
\end{equation}
It turns out that a simultaneous eigenstate of the mutually commuting 
transposition class-sum operator set 
$\{\bs{\cal C}_2[{\cal S}_n], \bs{\cal C}_2[{\cal S}_{n-1}],\ldots,
\bs{\cal C}_2[{\cal S}_2]\}$ has definite symmetry with respect to the
group-subgroup chain 
${\cal S}_n \supset {\cal S}_{n-1} \supset \ldots \supset {\cal S}_2
\supset {\cal S}_1$. Accordingly, NKG formulated a recursive method
that starts with one-particle states and constructs symmetrized $A$-particle
states through the following procedure:
\begin{itemize}
\item[a.] Add another particle (let say a total of $n$).
\item[b.] Identify the subspaces invariant under the action of the permutation
         group ${\cal S}_n$. Create the appropriate basis.
\item[c.] In each subspace, evaluate the matrix elements of
         the transposition class-sum operator
         $\bs{\cal C}_2[{\cal S}_n]$, and diagonalize it.
\item[d.] The ${\cal S}_n$ 
         symmetry of the resulting states is unambiguously identified 
         through the eigenvalues of $\bs{\cal C}_2[{\cal S}_n]$. 
         The eigenvectors of the $\bs{\cal C}_2[{\cal S}_n]$ matrix
         are the transformation coefficients from the old basis (c.) 
         to the new
         symmetrized one, these are the coefficients of fractional parentage.
\item[e.] Ensure phase consistency (see reference \cite{Akiva89a} for 
          a detailed account of this delicate point).
\item[f.] If $n< A$ repeat the procedure.
\end{itemize}
When this procedure is done, the original states are transformed
step by step into the desired symmetrized $A$-particle states.

In the case under consideration, 
the $n$-particle states are labeled by a set of good quantum 
numbers consisting of the total spin $S_n$ and its projection $S^z_n$ (which we
shall suppress), the color projection $G_n=(G^1_n, G^2_n)$ and the set of 
Young 
diagrams $\Gamma_n,\Gamma_{n-1},\ldots ,\Gamma_3,\Gamma_2,\Gamma_1$ which
is equivalent to the Yamanouchi symbol $Y_n$ \cite{Ham62}.
Using these labels one can construct a complete set of states labeled by
$| S_n G_n Y_n \beta_n \ket$, where $\beta_n$ is an additional
label that removes the remaining degeneracies.
For single particle, there is no degeneracy and one sees that
\begin{equation}
  | S_1 G_1 \Gamma_1 \ket = | \ratio{1}{2} g_1  \ket \,,
\end{equation}
where $S_1=\frac{1}{2}$ and $G_1=g_1$ or equivalently 
$(G^1_1,G^2_1)=(g^1_1, g^2_1)$.
To obtain a two-particle state labeled by the sequence of Young diagrams
$\Gamma_2, \Gamma_1$, 
we first couple the spin states, 
and introduce the notation
\begin{equation}
|(S_1 G_1 \Gamma_1 ; g_2) S_2 G_2 \ket =
 [ | S_1 G_1 \Gamma_1 \ket | \ratio{1}{2} g_2 \ket]^{S_2}
\end{equation}
to denote $2$-particle states with total spin $S_2$ and total color projection
$G_2=G_1+g_2$.
Here, and in what follows, we have used the short hand notation 
$G_2=G_1+g_2$ for 
$(G^1_2,G^2_2)=(G^1_1+g^1_2,G^2_1+g^2_2)$.
The permutation operators transform these states within an invariant subspace,
each of which is characterized by the total spin $S_2$ and the total
color projection $G_2=(G^1_2, G^2_2)$. 
The linear combinations which belong to well-defined irreps of the symmetry
group can be written in the form,
\begin{equation}
  | S_2 G_2 \Gamma_2 \Gamma_1 \beta_2 \ket = 
  \sum_{G_1 g_2 ; G_2=G_1+g_2}
  [S_1 G_1 \Gamma_1 g_2 |\} S_2 G_2 \Gamma_2 \beta_2]
  | (S_1 G_1 \Gamma_1 ;  g_2) S_2 G_2 \ket \;.
\end{equation}
Assume that we have constructed a $(n-1)-$particle states with well-defined
permutational symmetry. Following the $2-$particle example we first form
$n$-particle states with well-defined total spin and total color 
projection,
\begin{equation} \label{n_nonsym}
|(S_{n-1} G_{n-1} \Gamma_{n-1} \beta_{n-1}; g_n) S_n G_n \ket =
 [ | S_{n-1} G_{n-1} \Gamma_{n-1} \beta_{n-1} \ket 
   | \ratio{1}{2} g_n \ket]^{S_n} \,.
\end{equation}
Note that $G_n=G_{n-1}+g_n$.
The desired, symmetrized, $n$-particle states can now be written in the 
form
\begin{eqnarray}\label{cscfps_n}
  | S_n G_n Y_n \beta_n \ket &  = &
  \sum_{S_{n-1} \beta_{n-1} G_{n-1} g_n ; G_n=G_{n-1}+g_n} \cr &&
  [S_{n-1} G_{n-1} Y_{n-1} \beta_{n-1} g_n |\} S_n G_n Y_{n} \beta_n]
|(S_{n-1} G_{n-1} \Gamma_{n-1} \beta_{n-1}; g_n) S_n G_n \ket \,.
\end{eqnarray}
The transformation coefficients 
$[S_{n-1} G_{n-1} Y_{n-1} \beta_{n-1} g_n |\} S_n G_n Y_{n} \beta_n]$
are the color-spin coefficients of fractional parentage, cscfps. These
coefficients are the lines of an orthogonal transformation matrix. Therefore
they fulfill the orthogonality and completeness relations.

The evaluation of the matrix-elements of the transposition class-sum operator,
$\bs{\cal C}_2[{\cal S}_n]$, for the non-symmetrized $n$-particle states, 
Eq. (\ref{n_nonsym}), can be drastically simplified  
if we recast it in the form
\begin{equation}\label{C2SR}
 \bs{\cal C}_2[{\cal S}_n]=\bs{\cal C}_2[{\cal S}_{n-1}] 
                          +\sum_{i=1}^{n-1} (i,n)  \,.
\end{equation}
The first term on the right-hand side of Eq. (\ref{C2SR}) is diagonal
in $\Gamma_{n-1}$. Its explicit value can be calculated 
exploiting the following expression for the eigenvalues of 
$\bs{\cal C}_2[{\cal S}_n]$ \cite{partensky},
\begin{equation}
 \bs{\cal C}_2[\Gamma_n]=\frac{1}{2}\sum_{i} r_i(r_i-2i+1) \, ,
\end{equation}
where $r_i$ is the number of boxes in the $i$'th row of $\Gamma_n$
and the sum extends over all the rows.
The matrix elements of the second term between states
with well-defined ${\cal S}_{n-1}$ symmetry can be rewritten as \cite{NKG88},
\begin{eqnarray}
\lefteqn{ \sum_{i=1}^{n-1} 
 \bra \Gamma_{n-1} Y_{n-2} |(i,n)| \Gamma'_{n-1} Y'_{n-2} \ket} && \cr &&
 = \delta_{\Gamma_{n-1}, \Gamma'_{n-1}}\delta_{Y_{n-2}, Y'_{n-2}}
 \frac{n-1}{|\Gamma_{n-1}|}\sum_{Y''_{n-2}\in \Gamma_{n-1}}
 \bra \Gamma_{n-1} Y''_{n-2} |(n-1,n)| \Gamma_{n-1} Y''_{n-2} \ket \,.
\end{eqnarray}
The $(n-1,n)$ matrix elements can be evaluated by expanding the 
$(n-1)-$particle states in terms of the cscfps calculated previously
and by recoupling the spin states in different order,
\begin{eqnarray}\label{nnm1}
\lefteqn{
 \bra (S_{n-1} G_{n-1} \Gamma_{n-1} \beta_{n-1}; g_n) S_n G_n |
  (n-1,n)
  |(S'_{n-1} G'_{n-1} \Gamma_{n-1} \beta'_{n-1}; g'_n) S_n G_n \ket } &&
\cr
 & = &
  \sum_{S_{n-2} G_{n-2} \beta_{n-2} g_{n-1} g'_{n-1}} 
  [S_{n-2} G_{n-2} Y_{n-2} \beta_{n-2} g_{n-1} |\} 
   S_{n-1} G_{n-1} Y_{n-1} \beta_{n-1}] \cr & &
  [S_{n-2} G_{n-2} Y_{n-2} \beta_{n-2} g'_{n-1} |\} 
   S'_{n-1} G'_{n-1} Y_{n-1} \beta'_{n-1}]
  \cr && 
  \times \sum_{S_{n,n-1}} 
       \sixj{\ratio{1}{2}}{\ratio{1}{2}}{S_{n,n-1}}{S_{n-2}}{S_n}{S_{n-1}}
       \sixj{\ratio{1}{2}}{\ratio{1}{2}}{S_{n,n-1}}{S_{n-2}}{S_n}{S'_{n-1}}
  \cr &&
(-1)^{S_n+S_{n-2}} 
       (2 S_{n,n-1}+1)
       \sqrt{(2 S_{n-1}+1)(2 S'_{n-1}+1)} 
  \cr &&
       \bra S_{n,n-1} g_{n-1} g_n |(n-1,n)| S_{n,n-1} g'_{n-1} g'_n \ket\,.
\end{eqnarray}
Since the transposition $(n-1,n)$ commutes with the $(n-2)-$particle operators
the matrix element is diagonal in the $(n-2)-$particle states. The matrix
element on the right-hand side of Eq. (\ref{nnm1}) is the matrix element
between two-particle states after we have decoupled them from the rest of 
the system. This matrix element can easily be evaluated to yield,
\begin{equation}
   \bra S_{n,n-1} g_{n-1} g_n |(n-1,n)| S_{n,n-1} g'_{n-1} g'_n \ket
   = -(-1)^{S_{n,n-1}} \delta_{g_{n-1},g'_n} \delta_{g'_{n-1},g_n} \, .
\end{equation}
\subsection{$SU(3)$ singlet states and $C-$parity}

In the previous subsection we outlined a method to construct color--spin
states with well
defined permutational symmetry $Y_A$, spin quantum numbers $(S_A S^z_A)$,
and color projection $G_A$. Still, the $SU(3)$ color symmetry and 
$C-$parity of these
states is not established, since a state with the same color projection
can belong to numerous irreps of $SU(3)$. For example the color
projection $(0,0)$ can belong to the singlet representation $[0]$ as well
as to the octet representation, $[8]$. 
The $SU(3)$ symmetry of the $A$-particle states can be established
using the quadratic Casimir operator, $\bs{C}_2[SU(3)]$. 
Any $SU(3)$ state is an eigenvector of $\bs{C}_2[SU(3)]$ with eigenvalue that
depends on the specific representation of the state. In particular the
states with eigenvalue $0$ are singlet $SU(3)$ states.
The $A$-particle $SU(3)$ generators
\begin{equation}
  \lambda^c_{a} = \sum_{i=1}^{A} {\lambda}^c_{a\,i}\;,
\end{equation}
are symmetric with respect to particle permutations. Consequently
the quadratic Casimir,
\begin{equation}
\bs{C}_2[SU(3)]=\ratio{1}{4} \vec{{\lambda}}^c \cdot \vec{{\lambda}}^c
    =\ratio{1}{4}\sum_{a=1}^{8} {\lambda}^c_a {\lambda}^c_a \;,
\end{equation}
commutes with the permutation group ${\cal S}_A$, and the states 
$\{| S_A G_A Y_A \beta_A \ket\}$ with good
quantum numbers $(S_A G_A Y_A)$ form an invariant subspace for 
$\bs{C}_2[SU(3)]$.
The calculation of the matrix-elements of the 
quadratic Casimir can be simplified if we rewrite it 
as a sum of one-particle and two-particle terms
\begin{equation} \label{c2_su3_A}
\bs{C}_2[SU(3)] =
     \ratio{1}{4} \sum_{i=1}^A \vec{{\lambda}}^c_i \cdot \vec{{\lambda}}^c_i
 +   \ratio{1}{4} \sum_{i\neq j}^A \vec{{\lambda}}^c_i \cdot \vec{{\lambda}}^c_j
 \,.
\end{equation}
The first sum on the right-hand side of Eq. (\ref{c2_su3_A}) 
is just a sum of the single--particle quadratic Casimir operator, and its
matrix-elements are just
\begin{equation}
\bra S_A G_A Y_A \beta_A | 
 \ratio{1}{4} \sum_{i=1}^A \vec{{\lambda}}^c_i \cdot \vec{{\lambda}}^c_i
| S_A G_A Y_A \beta'_A \ket\  = \delta_{\beta_A, \beta'_A} \frac{4}{3}A \,,
\end{equation}
since the eigenvalue of the quadratic Casimir for both representations
$[3]$ and $[\bar 3]$ equals $4/3$.
Due to Schur's Lemma the matrix elements of $\bs{C}_2[SU(3)]$ are 
independent of the particular permutational symmetry state $Y_{A-1}$.
Exploiting this observation we can recast the second sum on the right-hand 
side of  Eq. (\ref{c2_su3_A}) into the form
\begin{eqnarray}
\lefteqn{ \bra S_A G_A Y_A \beta_A | 
  \ratio{1}{4} \sum_{i \neq j}^A \vec{{\lambda}}^c_i \cdot \vec{{\lambda}}^c_j
 | S_A G_A Y_A \beta'_A \ket\  } \cr 
 & = & \frac{A(A-1)}{2}\frac{1}{|\Gamma_A|}\sum_{Y''_{A-1}\in \Gamma_A}
    \ratio{1}{2} \bra S_A G_A \Gamma_A Y''_{A-1} \beta_A | 
      \vec{{\lambda}}^c_{A-1} \cdot \vec{{\lambda}}^c_A
    | S_A G_A \Gamma_A Y''_{A-1} \beta'_A \ket\  \,.
\end{eqnarray}
The $\vec{{\lambda}}^c_{A-1} \cdot \vec{{\lambda}}^c_A$ matrix elements
can be evaluated by expanding the $A$-particle states in terms of
the cscfps, Eq. (\ref{cscfps_n}), to decouple the last two particles, 
\begin{eqnarray}\label{c2b}
\lefteqn{ \bra S_A G_A \Gamma_A Y_{A-1} \beta_A | 
      \vec{{\lambda}}^c_{A-1} \cdot \vec{{\lambda}}^c_A
    | S_A G_A \Gamma_A Y_{A-1} \beta'_A \ket\ } \cr
   & = & \sum_{S_{A-1} G_{A-1} G'_{A-1} \beta_{A-1} \beta'_{A-1} g_A g'_A}
      \sum_{S_{A-2} G_{A-2} \beta_{A-2} g_{A-1} g'_{A-1}}
   \cr && \times
  [S_{A-1}  G_{A-1}  Y_{A-1} \beta_{A-1}  g_A  |\} S_A G_A Y_{A} \beta_A ]
  [S'_{A-1} G'_{A-1} Y_{A-1} \beta'_{A-1} g'_A |\} S_A G_A Y_{A} \beta'_A]
  \cr && \times
  [S_{A-2}  G_{A-2}  Y_{A-2} \beta_{A-2}  g_{A-1} 
                             |\} S_{A-1} G_{A-1}  Y_{{A-1}} \beta_{A-1}]
  \cr && \times
  [S_{A-2}  G_{A-2}  Y_{A-2} \beta_{A-2}  g'_{A-1} 
                             |\} S_{A-1} G'_{A-1} Y_{{A-1}} \beta'_{A-1}]
  \cr && \times
  \bra g_A g_{A-1} | \vec{{\lambda}}^c_{A-1} \cdot \vec{{\lambda}}^c_A
  | g'_A g'_{A-1} \ket \;.
\end{eqnarray}
The last term on the right-hand side of Eq. (\ref{c2b}) is the sum
\begin{equation}
  \bra g_A g_{A-1} | \vec{{\lambda}}^c_{A-1} \cdot \vec{{\lambda}}^c_A
  | g'_A g'_{A-1} \ket =
  \sum_a   \bra g_{A-1} | {{\lambda}}^c_{a\,A-1} | g'_{A-1} \ket
           \bra g_A     | {{\lambda}}^c_{a\,A  } | g'_A     \ket \;,
\end{equation}
which can be easily evaluated using the explicit form of the 
$SU(3)$ generators for the $[3]$ and $[\bar 3]$ irreps, Eq. (\ref{su3_33bar_states}).

Turning now to establish states with good $C-$parity, we note that the 
$C-$parity operator, $C$, commutes with ${\cal S}_A$, the spin operator 
and the quadratic Casimir of $SU(3)$. Therefore the states
$\{| S_A G_A Y_A \beta_A \ket\}$ with good
quantum numbers $(S_A G_A Y_A)$ that belong to the singlet irrep of $SU(3)$ 
form an invariant subspace for $C$.
In order to evaluate the matrix-elements of $C$ we first express our
states in terms of the spin coupled single-particle states, Eq. (\ref{chi_cs}).
Then we can replace each quark state by the appropriate antiquark state
and vice versa.
The eigenvectors of $C$ with eigenvalues $c=\pm1$ that belong to
the subspace of $\bs{C}_2[SU(3)]$ eigenstates with eigenvalue '0' are the
desired physical color-spin basis functions.

\section{A constituent quark model}
\label{model}

The charmonium spectra has been thoroughly studied using constituent quark models 
for more than 25 years. As an almost pure coulombic system, models based on coulomb plus 
confinement interactions have been used to describe most of its spectroscopic properties \cite{Eic78}.
To address the study of four-quark states we make use of a
standard constituent quark model inspired in these works and originally applied to the study
of the nonstrange baryon spectra and the baryon-baryon interaction \cite{Rep05}. 
This model has been generalized to all flavor sectors giving a reasonable description
of the meson spectra \cite{Vij03}, the baryon spectra \cite{Vij05}, 
and the scalar mesons once four-quark configurations were included \cite{Vij03b}.
Within this model hadrons are described as clusters of constituent (massive) quarks.
This is based on the assumption that the light-quark constituent quark mass appears because 
of the spontaneous breaking of the original $SU(3)_{L}\otimes SU(3)_{R}$ 
chiral symmetry at some momentum scale. 
In this domain of momenta, quarks interact through 
Goldstone boson exchange potentials. For the particular case of heavy
quarks, chiral symmetry is explicitly broken and therefore the boson exchanges do not contribute to 
the $qq$ interaction. Explicit expression of the boson exchange
interacting potentials and a more detailed discussion can be found in Ref. \cite{Vij03}.

Beyond the chiral symmetry breaking scale one expects the dynamics
being governed by QCD perturbative effects. They are taken into account
through the one-gluon-exchange (OGE) potential \cite{ruju}. The
nonrelativistic reduction of the OGE diagram in QCD for point-like quarks
presents a contact term that, when not treated perturbatively, leads to
collapse \cite{BHA80}. This is why one maintains the structure of the OGE, but
the $\delta$ function is regularized in a suitable way. This regularization is
justified based on the finite size of the constituent quarks and should be
therefore flavor dependent \cite{YYYY}. As a consequence, the OGE reads, 
\begin{equation} 
\label{OGE}
V_{OGE}(\vec{r}_{ij}) ={\frac{1}{4}}\alpha
_{s}\,(\vec{\lambda ^{c}} _{i}\cdot \vec{\lambda^{c}}_{j})\,\left\{
{\frac{1}{r_{ij}}}-{\frac{1}{6m_{i}m_{j}}}\vec{\sigma}_{i}\cdot
\vec{\sigma}_{j} \,{\frac{{e^{-r_{ij}/r_{0}(\mu )}}}{r_{ij}\,
r_0^2(\mu)}}\right\}  \, ,
\end{equation}
where $\lambda^{c}$ are the $SU(3)$ color matrices, $\alpha_s$ is the quark-gluon coupling constant,
and $r_0(\mu)=\hat r_0 \mu_{nn}/\mu_{ij}$, where $\mu_{ij}$ is the reduced mass
of the interacting quarks $ij$ ($n$ stands for the light quarks $u$ and $d$)
and $\hat r_0$ is a parameter to be determined from the data.

The strong coupling constant, taken to be constant for each flavor sector, has
to be scale-dependent when describing different flavor sectors \cite{Tita95}. Such
an effective scale dependence has been related to the typical momentum scale of
each flavor sector and assimilated to the reduced mass of the system
\cite{Halz93}. We use a strong coupling constant given by,
\begin{equation}
\alpha_s(\mu_{ij})={\frac{\alpha_0}{\ln\{(\mu_{ij}^2+\mu_0^2)/\gamma_0^2\}}} \, ,
\end{equation}
where $\alpha_0$, $\mu_0$ and $\gamma_0$ are parameters fitted within a
global description of the meson spectra~\cite{Vij03}.

Finally, any model imitating QCD should incorporate 
confinement. Lattice calculations in the quenched
approximation derived, for heavy quarks, a confining interaction linearly
dependent on the interquark distance. The consideration of sea quarks apart
from valence quarks (unquenched approximation) suggest a screening effect on
the potential when increasing the interquark distance \cite{Bal01}. Creation of light-quark
pairs out of vacuum in between the quarks becomes energetically preferable
resulting in a complete screening of quark color charges at large distances.
String breaking has been definitively confirmed
through lattice calculations \cite{SESAM} in coincidence with the quite rapid
crossover from a linear rising to a flat potential well
established in SU(2) Yang-Mills theories \cite{este}.
A screened potential simulating these results can be written as
\begin{equation}
\label{confi}
V_{CON}(\vec{r}_{ij})=-a_{c}\,(1-e^{-\mu_c\,r_{ij}})(\vec{\lambda^c}_{i}\cdot \vec{ \lambda^c}_{j})\,.
\end{equation}
At short distances this potential presents a linear behavior with an 
effective confinement strength $a=a_c \, \mu_c \, (\vec{\lambda^c}_i \cdot \vec{\lambda^c}_j)$, while
it becomes constant at large distances. 
Such screened confining potentials provide with an explanation to the
missing state problem in the baryon spectra \cite{miss}, improve the
description of the heavy meson spectra \cite{Vij04}, and justify the 
deviation of the meson Regge trajectories from the linear behavior
for higher angular momentum states \cite{golr}.

\section{Results and discussion}
\label{results}

The results for the $cc\bar c\bar c$ states have been obtained
in the framework of the hyperspherical harmonic formalism 
explained above up to the maximum value of $K$ within our computational
capabilities, being $K_{max}=20$ for positive parity states and 
$K_{max}=21$ for negative parity states with the exception 
of $J^{PC}=1^{+-}$ ($K_{max}=18$), $2^{++}$ and $2^{+-}$ ($K_{max}=22$) 
and $2^{--}$ ($K_{max}=23$). The two-body problem has been solved 
using the Numerov algorithm (see Ref. \cite{Vij03} for details) and all
the model parameters fixed in the description of the $q\bar q$
spectra \cite{Vij03,Vij03b}. Since the only
relevant two-meson decay thresholds for these four-quark systems are those formed by
two $c\bar c$ mesons, we summarize in Table \ref{t1} the results obtained for
the charmonium spectrum in Ref. \cite{Vij03} compared with the experimental data 
quoted by the Particle Data Group (PDG) \cite{Pdg05}. It can be observed that
the energy difference in the $P-$wave $c \bar c$ multiplets due to the non-central terms 
is less than 50 MeV.
Although these terms are known to play an important role in the description of the 
light $q\bar q$ states, their contribution becomes smaller as the mass of the heavy 
quark increases \cite{Vij03}. This allows to neglect them on the four-body sector, because 
they would only provide with a fine tune of the final results but making the solution of the four-body 
problem much more involved and time consuming.

To analyze the stability of these systems against dissociation through strong decay,
$cc\bar c\bar c\to M_1(c\bar c)+M_2(c\bar c)$, parity ($P$), $C-$parity
($C$), and total angular momentum ($J$) must be preserved. Assuming that the
decay takes place in a relative $S-$wave for the final state,
$J_{4q}=J_{M_1}\otimes J_{M_2}$ and $P_{4q}=P_{M_1}\cdot P_{M_2}$. Since the final
states are also eigenstates of $C-$parity one has $C_{4q}=C_{M_1}\cdot C_{M_2}$. 
Once the meson masses have been obtained, the corresponding thresholds can be 
computed simply adding the mesons masses $M_1$ and $M_2$, $T(M_1,M_2)=M_1+M_2$. 
In Table \ref{t2} we indicate the lowest two-meson decay threshold for each set of
quantum numbers. Four-quark states will be stable under strong 
interaction if their total energy lies below all possible, and allowed, two-meson 
thresholds. It is useful to define
\begin{equation}
\label{delta}
\Delta=M(q_1q_2\bar q_3\bar q_4)-T(M_1,M_2)\,,
\end{equation}
in such a way that if $\Delta>0$ the four-quark system will fall
apart into two mesons, while $\Delta<0$ will indicate that such strong decay is
forbidden and therefore the decay, if
allowed, must be weak or electromagnetic, being its width much narrower.

\begin{figure}
\caption{Energy of the $0^{++}$ (circles) and $0^{-+}$
(crosses) four-quark states as a function of $K$. The solid line corresponds to the
extrapolation used for the $0^{++}$ case while the long-dashed one corresponds
to the $0^{-+}$ state. The thresholds for each state are denoted by a dashed 
line ($0^{++}$) and a dotted-dashed line ($0^{-+}$).}
\label{f1}
\mbox{\epsfxsize=150mm\epsffile{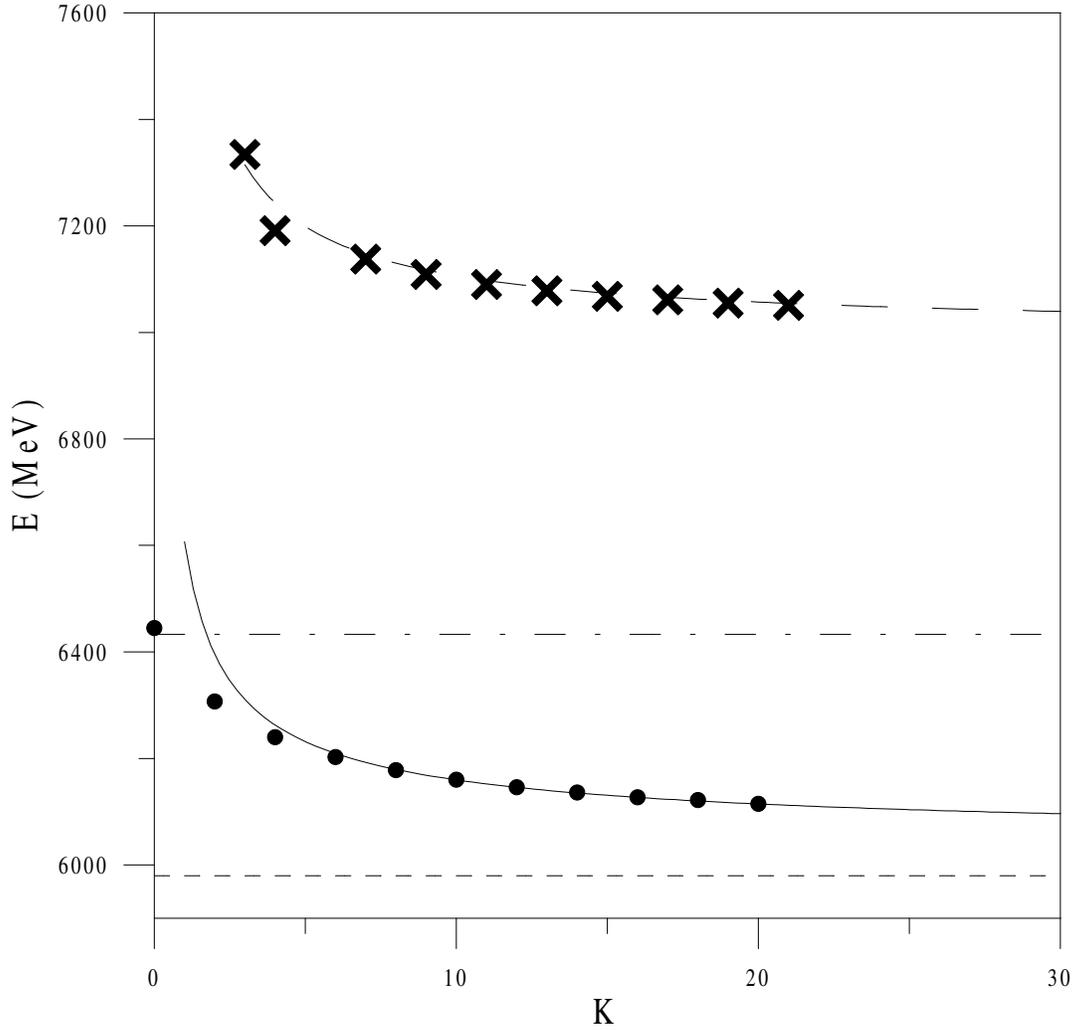}}
\end{figure}

Let us first of all analyze the convergence of the expansion in terms of hyperspherical
harmonics. We show in Fig. \ref{f1} the variation of the energy of the $J^{PC}=0^{++}$ and $0^{-+}$ cases 
with $K$, this behavior being similar for the other quantum numbers. From this figure it can be seen 
that the convergence is slow, and the effective potential techniques \cite{Bar00} are unable to improve it. In order 
to obtain a more adequate value for the energy we have extrapolated it according 
to the expression
\begin{equation}
\label{extra}
E(K)=E(K=\infty)+{\frac{a}{K^b}}\,,
\end{equation}
where $E(K=\infty)$, $a$ and $b$ are fitted parameters. In Table \ref{t2b}
we show the values obtained for $E(K=\infty)$ for the two states shown in Fig. \ref{f1} 
as a function of the fitting range $(K_0,K_f)$.  It can be checked that the values 
obtained for 
$E(K=\infty)$ are stable within $\pm$10 MeV for $K_0\geq10$.
The values obtained for the $b$ parameter are very similar for all possible
quantum numbers, being in all cases in the range $0.75-0.95$, corresponding the
lower limit to the positive parity states and the upper one to the negative
parity states. The origin of the difficulties to obtain a better convergence can be traced 
back to two different aspects of these systems. On the one hand, 
the HH formalism is better suited to deal with bound states, however, in this particular 
case we will show that most of the four-quark states are above the corresponding two-meson threshold. On the other 
hand, the color structure of a four-quark system is more involved than that of $q\bar q$ or $qqq$ systems \cite{Vij03b}, 
with two different possibilities to obtain a color singlet. This 
makes possible the existence of repulsive contributions due to the color operator of the interacting potential, 
Eqs. (\ref{OGE}) and (\ref{confi}). Although the total interaction is always confining, these deconfining terms make
the convergence much slower.

We have studied all possible $J^{PC}$ quantum numbers with $L=0$. We show in
Table \ref{t3} the results obtained for $K=K_{max}$, 
the maximum value of $K$ calculated, 
the results obtained using the extrapolation of Eq. (\ref{extra}), the corresponding threshold for each 
set of quantum numbers, and the value of $\Delta$, Eq. (\ref{delta}).
A first glance to the results shows that there are two sets of quantum numbers,
$J^{PC}=0^{+-}$ and $J^{PC}=2^{+-}$, where the four-quark configuration is
clearly below the corresponding two-meson threshold for dissociation. The
$J^{PC}=1^{+-}$ and $J^{PC}=2^{++}$ states are very close to the threshold and all the
other quantum numbers are far above the corresponding threshold.
Being all possible strong decays forbidden they should be narrow states with typical widths of 
the order of a few MeV.
It is also interesting to note that the quantum numbers $0^{+-}$ and $2^{+-}$ 
correspond to exotic states, those whose quantum numbers cannot be obtained from a
$q\bar q$ configuration, and therefore, if experimentally observed, they would be easily 
distinguished as a clear signal of pure non$-q\bar q$ states.

The energies obtained for the negative parity states are, as expected, much higher than those of
positive parity and therefore all of them are far above the corresponding thresholds, 
$\Delta(J^{PC}=J^{-C})>500$ MeV. Thus, they will fall apart immediately and therefore they should be very 
broad and difficult to detect. Let us also note that the experimental observation 
stating that a unit of angular momentum (parity change) in the $qqq$ and $q\bar q$ sector costs 
approximately 400$-$500 MeV of excitation energy in all flavor sectors (with some remarkable exceptions 
as the light scalar mesons \cite{Vij03b}) has an equivalent in the four-quark sector, being in this 
case the energy difference due to the parity change $E(J^{-C})-E(J^{+C})\approx800-900$ MeV. 

\begin{figure}
\caption{Variation of the energy of the four-quark states with exotic quantum numbers with the quark mass. 
The solid line corresponds to the quantum numbers $J^{PC}=0^{+-}$, dashed to
$2^{+-}$, long-dashed to $0^{--}$, and dashed-dotted to $1^{-+}$. The vertical
lines correspond to the light and charm quark masses, respectively.}
\vspace*{-4.0cm}
\mbox{\epsfxsize=150mm\epsffile{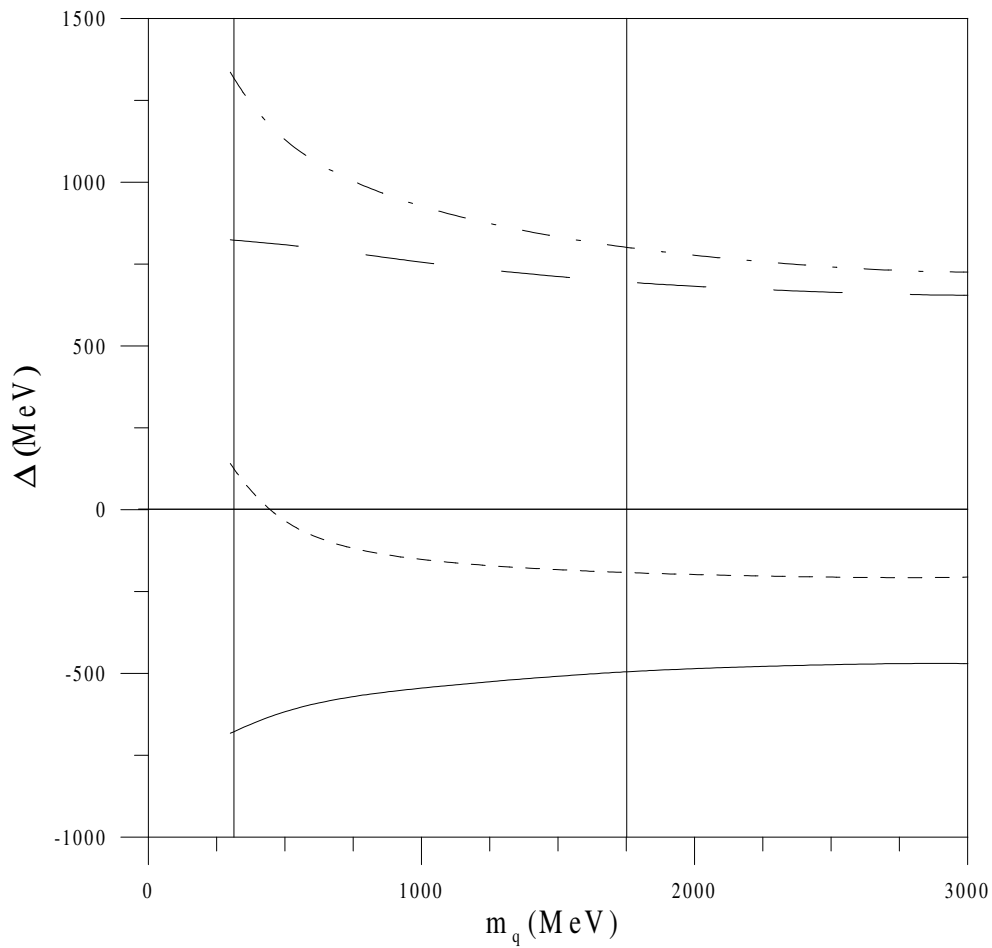}}
\label{f2}
\end{figure}

\begin{figure}
\caption{Variation of the energy of the non-exotic positive parity states with the quark mass. 
The grey box corresponds to the $\pm10$ MeV uncertainty in the extrapolation.
The solid line corresponds to the quantum numbers $J^{PC}=2^{++}$, dashed to
$1^{+-}$, long-dashed to $0^{++}$, and dashed-dotted to $1^{++}$. The vertical
lines correspond to the light and charm quark masses, respectively.}
\vspace*{-4.0cm}
\mbox{\epsfxsize=150mm\epsffile{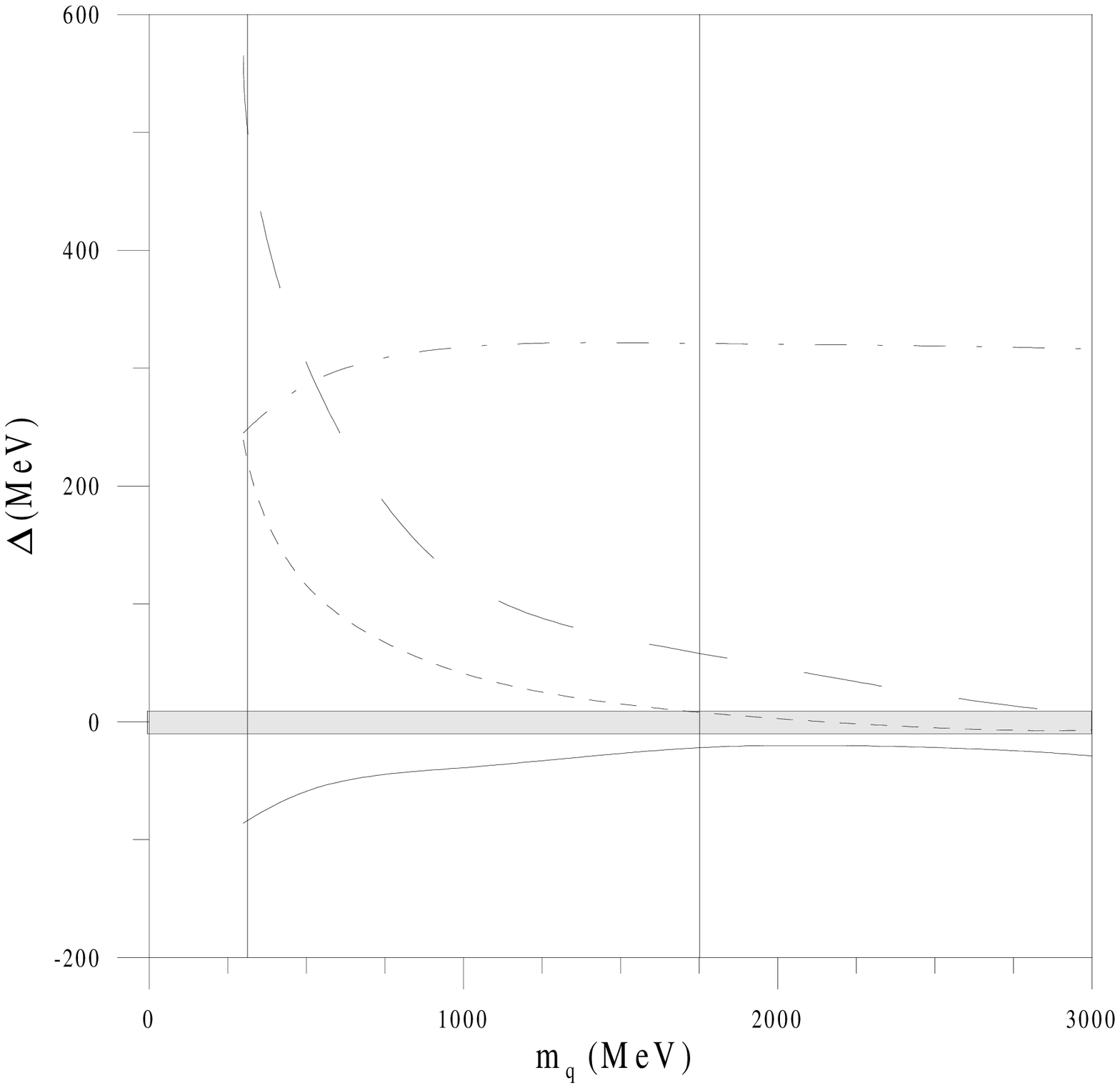}}
\label{f3}
\end{figure}

The existence of non$-q\bar q$ signals in the meson spectra has been 
the subject of an intensive debate \cite{Ams04}. 
To analyze whether the existence of bound states with exotic quantum numbers could be a characteristic 
feature of the heavy quark sector 
or it is also present in the light sector we show in Fig. \ref{f2} the value of $\Delta$ as a function of the quark mass
for all exotic quantum numbers and in Fig. \ref{f3} for the remaining positive parity states.
Since each quantum number has a different threshold these figures should be interpreted carefully.
One should notice that the value $\Delta=0$ in both figures 
corresponds to $M(q_1q_2\bar q_3\bar q_4)=T(M_1,M_2)$, and therefore, the fact that one state is 
below the others in these figures do not imply that its total mass would be smaller. 
Let us also note that since the heavy quarks are 
isoscalar states, the flavor wave function of the four heavy-quark
states will be completely symmetric with total isospin equal to zero.
Therefore, one should 
compare the results obtained in the light-quark case with a completely symmetric flavor 
wave function, i.e., the isotensor states.

In Fig. \ref{f3} we observe how one of the four-quark non-exotic positive
parity states, the $2^{++}$, becomes more bound when the quark mass is
decreased, $\Delta\approx-80$ MeV for $m_q=313$ MeV. The $1^{+-}$ and $0^{++}$
states, that were slightly above the threshold for $m_q=1752$ MeV, increase
their attraction when the quark mass is increased and only for masses above 3
GeV, close to the bottom quark mass, may be bound. With respect to the
exotic quantum numbers, we observe in Fig. \ref{f2} that the $0^{--}$ and
$1^{-+}$ are not bound for any value of the quark mass. The $2^{+-}$ state decreases its
binding when the quark mass diminishes and it becomes unbound for masses of the
order of 500 MeV, the strange quark mass.
Only the $0^{+-}$ four-quark state becomes more deeply bound when the constituent quark mass decreases, 
and therefore only one bound state with exotic quantum numbers would remain in the light-quark sector.

Taking the experimental mass for the threshold $T(M_1,M_2)$ in the light-quark case, one can estimate the energy region 
where these states can be found, being $M(0^{++})\approx 900$ MeV, $M(1^{+-})\approx 1200$ MeV,
$M(1^{++})\approx 1900$ MeV, and $M(2^{++})\approx 1500$ MeV. 
Note that the boson exchange potentials, that
as explained in Sect. II are present in the description of the light-quark sector, have not been considered in 
this analysis.  Although these terms do play a role in the description of the four-quark state total energy, 
it was shown in Ref. \cite{Vij03c} that their suppression will not alter the relative order of the four-quark states.
The four-quark states with quantum numbers
$0^{++}$ and $2^{++}$ have been analyzed with the complete constituent 
quark model including boson exchanges 
using a variational approach \cite{Vij03b}, obtaining 1004 MeV and 1500 MeV respectively, 
both in good agreement with our calculation and therefore giving confidence
on our approach neglecting the 
boson exchange interactions in this particular analysis. The states with exotic quantum numbers would be higher in energy,
being $M(0^{--})\approx M(1^{-+})\approx 2900$ MeV, $M(0^{+-})\approx 1800$ MeV and
$ M(2^{+-})\approx 2100$ MeV. 

There are experimental evidences for three states with exotic quantum numbers 
in the light-quark sector. Two of them are isovectors with quantum numbers $J^{PC}=1^{-+}$ named
$\pi_1(1400)$ and $\pi_1(1600)$, 
and one isotensor $J^{PC}=2^{++}$, the $X(1600)$ \cite{Pdg05}.
Based on the coupling of the $\pi_1(1400)$ to the $\eta\pi$ decay channel a possible four-quark structure has been 
suggested \cite{Chun02} while the strong $\eta'\pi$ coupling of the $\pi_1(1600)$ makes it a good candidate
for a pure hybrid state \cite{Iddi01}. The large mass obtained in our analysis 
for these quantum numbers makes doubtful 
the identification of any of them, and in particular of the $\pi_1(1400)$, with a pure four-quark state, although
a complete calculation with the proper flavor wave function is needed before
drawing a definitive conclusion \cite{Fut05}. An alternative explanation of the $\pi_1(1400)$ in terms of low-energy
rescattering effects has been proposed in Ref. \cite{Szc03} while recent reanalysis of experimental
data from E852 Collaboration with improved statistics show no evidence of any
exotic meson with a mass close to 1.6 GeV \cite{Dzi05}.
Concerning the $X(1600)$, being its experimental mass 1600$\pm$100 MeV, a possible tetraquark configuration
seems likely.

\section{Summary}
\label{summary}

In this work we have presented for the first time a generalization of the HH
formalism to study systems made of quarks and antiquarks of the same flavor. For
this purpose we made use of standard techniques widely used in nuclear physics
that have been adapted to include the color degree of freedom and to treat
explicitly the $C-$parity. This formalism opens the door to an exact study of
multiquark systems up to now described by means of different techniques, being
the variational methods the most standard ones. The particular color structure
of four-quark systems makes the convergence slow but attainable.

The formalism has been applied to study the existence of $L=0$ four-charm quark states.
For this purpose we made use of a well-established constituent quark model that
properly accounts for the charmed meson and baryon spectra. Our results suggest 
the possible existence of three four-quark bound states with quantum numbers $0^{+-}$,
$2^{+-}$ and $2^{++}$ and masses of the order of 6515, 6648, and 6216 MeV. The
two states with exotic quantum numbers, clearly below their 
corresponding two-meson threshold, should
present narrow widths and, if produced, may be easily detected. 

We have analyzed the variation of our results with the constituent quark mass. In the light-quark
case only the $0^{+-}$ and $2^{++}$ quantum numbers remain bound, being the
$0^{+-}$ the lowest one with an energy close to 1800 MeV.
The four-quark state with quantum numbers $1^{-+}$ lies around 2900 MeV, far from 
the experimental states $\pi_1(1400)$ and $\pi_1(1600)$, therefore supporting a 
hybrid configuration or a more complicated structure in terms of non-resonant 
structures. A possible description of the $X(1600)$ as a four-quark 
has also been justified.

The program we have started for and exact study of multiquark systems by means
of the HH formalism will be accomplished by implementing the possibility of
treating quarks of different masses. When this is done we will have at our
disposal a powerful method, imported from the nuclear physics, to study in an
exact way systems made of any number of quarks and antiquarks coupled to a
color singlet.

\section{Acknowledgments}
This work has been partially funded by Ministerio de Ciencia y 
Tecnolog\'{\i}a under Contract No. FPA2004-05616, by Junta de Castilla y Le\'{o}n under
Contract No. SA-104/04, and by Generalitat Valenciana under Contract No.
GV05/276.


\newpage

\begin{table}
\caption{Charmonium spectrum in MeV. Experimental data are taken
from PDG \protect\cite{Pdg05}.}
\label{t1}
\begin{center}
\begin{tabular}{|cccc|}
\hline
$(nL) \, J^{PC}$        & State                 & CQM   & Exp. \\
\hline
$(1S) \, 0^{-+}$        & $\eta_c(1S)$          & 2990  &2979.6$\pm1.2$ \\
$(1S) \, 1^{--}$        & $J/\psi(1S)$          & 3097  &3096.916$\pm0.011$ \\
$(1P) \, 0^{++}$        & $\chi_{c0}(1P)$       & 3443  &3415.19$\pm0.34$ \\
$(1P) \, 1^{++}$        & $\chi_{c1}(1P)$       & 3496  &3510.59$\pm0.10$ \\
$(1P) \, 2^{++}$        & $\chi_{c2}(1P)$       & 3525  &3556.26$\pm0.11$ \\
$(1P) \, 1^{+-}$        & $h_c(1P)$             & 3507  &3526.21$\pm$0.25 \\
$(2S) \, 0^{-+}$        & $\eta_c(2S)$          & 3627  &3654$\pm$10 \\
$(2S) \, 1^{--}$        & $\psi(2S)$            & 3685  &3686.093$\pm0.034$ \\
$(1D) \, 1^{--}$        & $\psi(3770)$          & 3776  &3770$\pm2.4$ \\ 
$(1D) \, 2^{--}$        & $\psi(3836)$          & 3790  &3836$\pm$13 \\ 
$(3S) \, 1^{--}$        & $\psi(4040)$          & 4050  &4040$\pm$10 \\ 
$(2D) \, 1^{--}$        & $\psi(4160)$          & 4104  &4159$\pm$20 \\
\hline
\end{tabular}
\end{center}
\end{table}  

\begin{table}
\caption{Lowest $S-$wave two-meson thresholds (MeV) for all $J^{PC}$ quantum numbers.}
\label{t2}
\begin{center}
\begin{tabular}{|ccc|}
\hline
$J^{PC}$  & $M_1M_2$                            & $T(M_1,M_2)$  \\
\hline
$ 0^{++}$ & $\eta_c(1S)\,\,\eta_c(1S)$          & 5980  \\ 
$ 0^{+-}$ & $\chi_{c1}(1P)\,\,h_c(1P)$          & 7003  \\
$ 1^{++}$ & $J/\psi(1S)\,\,J/\psi(1S)$          & 6194  \\
$ 1^{+-}$ & $J/\psi(1S)\,\,\eta_c(1S)$          & 6087  \\
$ 2^{++}$ & $J/\psi(1S)\,\,J/\psi(1S)$          & 6194  \\
$ 2^{+-}$ & $\eta_c(1S)\,\,\psi(3836)$          & 6780  \\
$ 0^{-+}$ & $\eta_c(1S)\,\,\chi_{c0}(1P)$       & 6433  \\
$ 0^{--}$ & $J/\psi(1S)\,\,\chi_{c1}(1P)$       & 6593  \\
$ 1^{-+}$ & $\eta_c(1S)\,\,\chi_{c1}(1P)$       & 6486  \\
$ 1^{--}$ & $\eta_c(1S)\,\,h_c(1P)$             & 6497  \\
$ 2^{-+}$ & $\eta_c(1S)\,\,\chi_{c2}(1P)$       & 6515  \\
$ 2^{--}$ & $J/\psi(1S)\,\,\chi_{c1}(1P)$       & 6593  \\
\hline
\end{tabular}
\end{center}
\end{table}

\begin{table}
\caption{$E(K=\infty)$ (MeV) as a function of the fitting range ($K_0,K_f$) for $J^{PC}=0^{++}$
and $J^{PC}=0^{-+}$.}
\label{t2b}
\begin{center}
\begin{tabular}{|c|c||c|c|}
\hline
($K_0,K_f$)     &       $E(K=\infty)[0^{++}]$   &       $(K_0,K_f)$     &       $E(K=\infty)[0^{-+}]$   \\
\hline
(2$,$20)        &       5831                                    &       (3$,$21)        &       7018    \\
(4$,$20)        &       5947                                    &       (5$,$21)        &       7007    \\
(6$,$20)        &       5997                                    &       (7$,$21)        &       7004    \\
(8$,$20)        &       6017                                    &       (9$,$21)        &       7003    \\
(10$,$20)       &       6030                                    &       (11$,$21)       &       7000    \\
(12$,$20)       &       6038                                    &       (13$,$21)       &       7000    \\
(14$,$20)       &       6041                                    &       (15$,$21)       &       6993    \\
(16$,$20)       &       6047                                    &       (17$,$21)       &       6990    \\
\hline
\end{tabular}
\end{center}

\end{table}  
\begin{table}
\caption{$cc\bar c\bar c$ masses obtained for the maximum value of $K$ computed, $E(K_{max})$, and using 
the extrapolation of Eq. (\ref{extra}), $E(K=\infty)$, compared with the corresponding threshold for each set of quantum numbers,
$T(M_1,M_2)$, as
given in Table \ref{t2}. The value of $\Delta$ for each state is also given.
All energies are in MeV.}
\label{t3}
\begin{center}
\begin{tabular}{|c|cc|c|c|}
\hline
$J^{PC}$  & $E(K_{max})$        &$E(K=\infty)$  &  $T(M_1,M_2)$ &       $\Delta$\\
\hline
$ 0^{++}$ &             6115            &       6038            & 5980                  &       +58     \\ 
$ 0^{+-}$ &             6606            &       6515            & 7003                  &       $-$488  \\
$ 1^{++}$ &             6609            &       6530            & 6194                  &       +336\\
$ 1^{+-}$ &             6176            &       6101            & 6087                  &       +14\\
$ 2^{++}$ &             6216            &       6172            & 6194                  &       $-$22\\
$ 2^{+-}$ &             6648            &       6586            & 6780                  &       $-$194\\
$ 0^{-+}$ &             7051            &       6993            & 6433                  &       +560\\
$ 0^{--}$ &             7362            &       7276            & 6593                  &       +683\\
$ 1^{-+}$ &             7363            &       7275            & 6486                  &       +789\\
$ 1^{--}$ &             7052            &       6998            & 6497                  &       +501\\
$ 2^{-+}$ &             7055            &       7002            & 6515                  &       +487\\
$ 2^{--}$ &             7357            &       7278            & 6593                  &       +685\\
\hline
\end{tabular}
\end{center}
\end{table}

\begin{thebibliography}{99}

\bibitem{Bel90} V.B. Belyaev, 
                {\it Lectures on the Theory of Few-body Systems}, (Springer-Verlag, Berlin, 1990).

\bibitem{Suz98} Y. Suzuki and K. Varga,
                {\it Stochastic Variational Approach to Quantum Mechanical Few-Body
                Problems}, (Springer, Berlin, 1998).

\bibitem{Nir9798} N. Barnea and A. Novoselsky, 
                        Ann. Phys. (N.\ Y.) {\bf 256}, 192 (1997); 
                        Phys. Rev. A {\bf 57}, 48 (1998).

\bibitem{Bac04} S. Bacca {\it et al.},
                        Phys. Lett. B {\bf 603}, 159 (2004).

\bibitem{Vij05} A. Valcarce, H. Garcilazo, and J. Vijande,
                Phys. Rev. C {\bf 72}, 025206 (2005).

\bibitem{Jaf77} R.L. Jaffe, 
                        Phys. Rev. D {\bf 15}, 267 (1977). 

\bibitem{Iwa76} Y. Iwasaki, 
                Prog. Theor. Phys. {\bf 54}, 492 (1975).

\bibitem{Ros76} C. Rosenzweig,
           Phys. Rev. Lett. {\bf 36}, 697 (1976);
                                B.G. Kenny, D.C. Peaslee and L.J. Tassie,
                Phys. Rev. D {\bf 14}, 302 (1976);
                                K.-T. Chao,
                Z. Phys. C {\bf 7}, 317 (1981).

\bibitem{Jaf04} R.L. Jaffe,
        Phys. Rep. {\bf 409}, 1 (2005) and references therein.

\bibitem{Ams04} C. Amsler and N.A. Tornqvist,
        Phys. Rep. {\bf 389}, 61 (2004) and references therein.  

\bibitem{Cho03} S.-K.Choi {\it et al.},         
                Phys. Rev.  Lett. {\bf 91}, 262001 (2003).

\bibitem{Aub05} B. Aubert {\it et al.},
                        Phys. Rev. Lett. {\bf 95}, 142001 (2005).

\bibitem{Aub03} B. Aubert {\it et al.},
           Phys. Rev. Lett. {\bf 90}, 242001 (2003);
                                D. Besson {\it et al.},
               Phys. Rev. D {\bf 68}, 032002 (2003);
                                K. Abe {\it et al.},
           Phys. Rev. D {\bf 69}, 112002 (2004).

\bibitem{Bar05} T. Barnes, F.E. Close, and H.J. Lipkin,
        Phys. Rev. D {\bf 68}, 054006 (2003);
                                A.P. Szczepaniak, 
                Phys. Lett. B {\bf 567}, 23 (2003);
                        K. Terasaki,
                Phys.  Rev. D {\bf 68}, 011501(R) (2003);  
                        H.Y. Cheng and W.S. Hou,
                Phys.  Lett. B {\bf 566}, 193 (2003);
                        Y.-Q. Chen and X.-Q. Li,
                Phys. Rev.  Lett. {\bf 93}, 232001 (2003);
                        T. Browder, S. Pakvasa, and A. Petrov,
                Phys. Lett. B {\bf 578}, 365 (2004);
                                J. Vijande, F. Fern\'andez, and A. Valcarce,
		Phys. Rev. D {\bf 73}, 034002 (2006);
                S.L. Olsen,
                Int. J. Mod. Phys. A {\bf 20}, 240 (2005);
                L. Maiani, F. Piccinini, A.D. Polosa, and V. Riquer,
                Phys. Rev. D {\bf 71}, 014028 (2005);
                L. Maiani, V. Riquer, F. Piccinini, and A.D. Polosa,
                Phys. Rev. D {\bf 72}, 031502 (2005).

\bibitem{Ade82} J.P. Ader, J.-M. Richard, and P. Taxil,
        Phys. Rev. D {\bf 25}, 2370 (1982);
                J.L. Ballot and J.-M. Richard,
        Phys. Lett. B {\bf 123}, 449 (1983);
                H.J. Lipkin,
        Phys. Lett. B {\bf 172}, 242 (1986);

\bibitem{Hel87} L. Heller and J.A. Tjon,
                Phys. Rev. D {\bf 32}, 755 (1985); {\it ibid} {\bf 35}, 969 (1987).

\bibitem{Car88} J. Carlson, L. Heller, and J.A. Tjon,
                Phys. Rev. D {\bf 37}, 744 (1988);
                A.V. Manohar and M.B. Wise,
                Nucl. Phys. B {\bf 399}, 17 (1993);
                S. Pepin, Fl. Stancu, M. Genovese, and J.-M.  Richard,
                Phys. Lett. B {\bf 393}, 119 (1997);

\bibitem{Sil93} B. Silvestre-Brac and C. Semay,
                Z. Phys. C {\bf 57}, 273 (1993); {\it ibid} {\bf 59}, 457 (1993); {\it ibid} {\bf 61}, 271 (1994);

\bibitem{Vij03c} J. Vijande, F. Fern\'andez, A. Valcarce, and B.  Silvestre-Brac,
                Eur. Phys. J. A {\bf 19}, 383 (2004).

\bibitem{Vij03b} J. Vijande, A. Valcarce, F. Fern\'andez, and B.  Silvestre-Brac, 
                Phys. Rev. D {\bf 72}, 034025 (2005).

\bibitem{Sil92} B. Silvestre-Brac,
                Phys. Rev. D {\bf 46}, 2179 (1992).

\bibitem{Llo04} R.J. Lloyd and J.P. Vary,
                Phys. Rev. D {\bf 70}, 014009 (2004).

\bibitem{Bar00} N. Barnea, W. Leidemann, and G. Orlandini,
                Phys. Rev. C {\bf 61}, 054001 (2004).
        
\bibitem{Vij03} J. Vijande, F. Fern\'andez, and A. Valcarce, 
                J. Phys. G {\bf 31}, 481 (2005).

\bibitem{NKG88} A. Novoselsky, J. Katriel and R. Gilmore, 
                        J. Math. Phys. {\bf 29}, 1368 (1988). 

\bibitem{BLO99} N. Barnea, W. Leidemann, and G. Orlandini, 
                        Nucl. Phys. A {\bf 650}, 427 (1999).

\bibitem{Akiva89a} A. Novoselsky and J. Katriel, 
                        Ann. Phys. (N. Y.) {\bf 196}, 135 (1989). 

\bibitem{Ham62} M. Hamermesh, 
                {\it Group Theory and its Application to Physical Problems}, (Addison-Wesley, Massachusetts, 1962).

\bibitem{partensky}  A. Partensky and C. Maguin,
                        J. Math. Phys. {\bf 19}, 511 (1978). 

\bibitem{Eic78} E. Eichten {\it et al.},
                Phys. Rev. D {\bf 17}, 3090 (1978); {\it ibid} D {\bf 21}, 313
                (E) (1980);
                R.K. Bhaduri, L.E. Cohler, and Y. Nogami,
                Nuovo Cim. A {\bf 65}, 376 (1981); 
                N. Isgur and G. Karl,
                Phys. Today {\bf 36}, 36 (1983).

\bibitem{Rep05} A. Valcarce, H. Garcilazo, F. Fern\'andez, and P. Gonz\'alez,
                Rep. Prog. Phys. {\bf 68}, 965 (2005) .

\bibitem{ruju} A. de R\'{u}jula, H. Georgi, and S.L. Glashow,         
                Phys.  Rev. D {\bf 12}, 147 (1975).  

\bibitem{BHA80} R.K. Bhaduri, L.E. Cohler, and Y. Nogami,
        Phys. Rev. Lett. {\bf 44}, 1369 (1980). 

\bibitem{YYYY} J. Weinstein and N. Isgur,
        Phys. Rev. D {\bf 27}, 588 (1983).

\bibitem{Tita95} S. Titard and F.J. Yndurain,         
                Phys. Rev. D {\bf 51}, 6348 (1995).

\bibitem{Halz93} F. Halzen, C. Olson, M.G. Olsson and M.L. Stong,
        Phys. Rev. D {\bf 47}, 3013 (1993).

\bibitem{Bal01} G.S. Bali,
               Phys. Rep. {\bf 343}, 1 (2001) and references therein.

\bibitem{SESAM}  SESAM Collaboration, G.S. Bali, H. Neff, T. D\"ussel, T.  Lippert, and K. Schilling,
        Phys. Rev. D {\bf 71}, 114513 (2005).

\bibitem{este} P.W. Stephenson,
          Nucl. Phys. B {\bf 550}, 427 (1999);
                  F. Knechtli and R. Sommer,
          Nucl. Phys. B {\bf 590}, 309 (2000);
                  O. Philipsen and H. Wittig,
          Phys. Lett. B {\bf 451}, 146 (1999);
                      P. de Forcrand and O. Philipsen,
          Phys. Lett. B {\bf 475}, 280 (2000).

\bibitem{miss} J. Vijande, P.  Gonz\'alez, H. Garcilazo, and A.  Valcarce,
              Phys.  Rev.  D {\bf 69}, 074019 (2004).

\bibitem{Vij04}P. Gonz\'alez, A. Valcarce, H. Garcilazo, J. Vijande
          Phys.  Rev.  D {\bf 68}, 034007 (2003). 

\bibitem{golr} M.M.  Brisudov\'a, L.  Burakovsky, and T.  Goldman,
              Phys.  Rev.  D {\bf 61}, 054013 (2000).  

\bibitem{Pdg05} S. Eidelman {\it et al.}, 
                        Phys. Lett. B {\bf 592}, 1 (2004).

\bibitem{Chun02} S.U. Chung, E. Klempt, and J.G.  Korner,
                        Eur. Phys. J. A {\bf 15}, 539 (2002).

\bibitem{Iddi01} F. Iddir and A.S. Safir,
                        Phys. Lett. B {\bf 507}, 183 (2001).

\bibitem{Fut05} N. Barnea, J. Vijande, and A. Valcarce,
                        in preparation.

\bibitem{Szc03} A.P. Szczepaniak, M. Swat, A.R. Dzierba, and S. Teige,
        Phys. Rev. Lett. {\bf 91}, 092002 (2003). 

\bibitem{Dzi05} A.R. Dzierba {\it et al.}, hep-ex/0510068.

\end{thebibliography}
\end{document}